\def\year{2021}\relax

\documentclass[letterpaper]{article} 
\usepackage{aaai21}  
\usepackage{times}  
\usepackage{helvet} 
\usepackage{courier}  
\usepackage[hyphens]{url}  
\usepackage{graphicx} 
\usepackage{natbib}  
\usepackage{caption} 
\usepackage{amsmath}
\usepackage{tikz}
\usepackage{subcaption}
\usepackage{adjustbox}
\title{Measuring Human Adaptation to AI in Decision Making:\\
Application to Evaluate Changes after AlphaGo}

\author{
    Minkyu Shin\textsuperscript{\rm 1},
        Jin Kim\textsuperscript{\rm 1}, 
    Minkyung Kim\textsuperscript{\rm 2} \\

}
\affiliations {
    \textsuperscript{\rm 1}Yale University\\
    \textsuperscript{\rm 2}University of North Carolina at Chapel Hill\\
    minkyu.shin@yale.edu,
    jin.m.kim@yale.edu,
    Minkyung\_Kim@kenan-flagler.unc.edu 
    
}
\begin{document}

\maketitle

\begin{abstract}
Across a growing number of domains, human experts are expected to learn from and adapt to AI with superior decision making abilities. But how can we quantify such human adaptation to AI? We develop a simple measure of human adaptation to AI and test its usefulness in two studies. In Study 1, we analyze 1.3 million move decisions made by professional Go players and find that a positive form of adaptation to AI (learning) occurred after the players could observe the \textit{reasoning processes} of AI, rather than mere \textit{actions} of AI. These findings based on our measure highlight the importance of explainability for human learning from AI. In Study 2, we test whether our measure is sufficiently sensitive to capture a negative form of adaptation to AI (cheating aided by AI), which occurred in a match between professional Go players. We discuss our measure's applications in domains other than Go, especially in domains in which AI's decision making ability will likely surpass that of human experts.
\end{abstract}

\section{Introduction}

As Artificial Intelligence (AI) technology advances, AI will have an ever greater impact on human decision making. But how big will this impact be? Will humans adapt to AI by learning from it and making better decisions themselves? If so, how quickly or slowly will humans adapt to AI? Answering these questions will become easier with better methods of measuring AI’s impact on human decision making. To this end, we propose an intuitive and objective measure of human adaptation to AI: the \textit{Human-AI Gap}. By comparing the quality of human decisions to that of superhuman AI decisions, we can quantify the \textit{Human-AI Gap} and use this measure to understand human adaptation to AI. We test our measure in an empirical setting in the game of Go\footnote{Go is a board game between two players who take turns placing ``stones" of their color (black or white) on a 19$\times$19 grid of lines. The game's objective is to surround a larger territory on the board than the opponent by completely enclosing it with one's stones.} and find that human players adapted to AI. 

In proposing our measure of human adaptation to AI, we examine the game of Go for two reasons. First, it is one of the first domains in which AI achieved superhuman performance in a complex decision making problem \cite{silver2016mastering}. Superhuman performance is a necessary condition for measuring AI’s impact on human decision making, not only because it encourages humans to learn from AI, but also because it provides an objective standard to which the quality of human decision making can be compared. This latter point becomes obvious when we compare a superhuman AI to a par-human AI. A superhuman AI makes decisions of superhuman quality, and it can evaluate how inferior human decisions are compared to its own decisions. In contrast, a par-human AI makes decisions of human-level quality, and it cannot evaluate the quality of human decisions with any authority, since it is no better than humans. Likewise, a superhuman AI can measure \textit{improvement} in human decision quality by observing a closing gap between human and AI decision quality, while a par-human AI would be unable to appreciate an increase in human decision quality. Now that AI is expected to surpass humans in many other decision making domains, our measure, which takes advantage of a superhuman AI, can be useful in domains beyond Go.

Another reason we examine the game of Go is that a game is an effective setting to test how humans interact with AI and adapt their decision making. The goal of a game is usually well-defined and human players choose various actions to achieve the goal. Those actions, or any decisions, and their resulting changes in the environment are  recorded in a database. Using these unique features of a game, researchers have studied various aspects of human decision making, from error correction to skill acquisition \cite{biswas2015measuring,regan2014human,stafford2014tracing,strittmatter2020life,tsividis2017human}. In addition, compared to other games, Go presents arguably the most complex task, which explains why AlphaGo defeating a top human expert was seen as a major breakthrough for artificial intelligence. If our measure of human adaptation is useful in a complex, computationally challenging domain as the game of Go, we expect it to be useful in other settings where decision making process is also complex.

We test the usefulness of our measure in two empirical studies. First, we examine the impact of AI on human experts' decision quality. Using historical data on human decisions before and after the emergence of AI such as AlphaGo, we investigate whether and when humans learn to make better decisions like AI. Although additional information provided by AI could be beneficial to human players, the black-box nature of AI's decisions may have hindered human adaptation or generated misinterpretation. Indeed, our results suggest that merely observing AI's \textit{actions} may not bring a meaningful improvement in human decision making. Observing AI's \textit{reasoning processes}, however, does seem to improve human decision making. Second, we also study whether our measure can detect a cheating behavior. Because AI outperforms human experts in the game of Go, it would be tempting for human experts to refer to AI's decisions in a match between humans. Not surprisingly, there have been reports of cheating using an AI program not only in professional Go matches but also in chess matches, particularly as more matches are held online during the pandemic. We show that our measure can detect an instance of cheating that has actually occurred.

Our measure has broad applications beyond Go. A natural extension is to measure users' performance in other games such as chess. We can use the measure in many other domains in which final decisions of high stakes are made by humans despite a presence of superhuman AI (e.g., medicine). Moreover, we can use the measure to investigate whether some people adapt to AI at a faster or slower rate than others, and whether AI widen a social gap between people.

\section{A Measure of Human Adaptations to AI}

Quantifying the quality of decisions has been made possible by the recent developments in Reinforcement Learning (henceforth RL). We take advantage of outputs from RL and apply them to study human adaptation to AI. This approach is reasonable since RL has been studied not just for developing an effective AI, but also for explaining human skill learning. For example, many computer scientists have made RL-based AI programs that solve complex decision problems ranging from playing complex board games to scheduling educational activities \cite{bassen2020reinforcement,mnih2013playing,nazari2018reinforcement,silver2017mastering}. Cognitive scientists also have used RL to study how humans learn new skills \cite{fiorillo2003discrete,holroyd2002neural,niv2009reinforcement,waelti2001dopamine}. Recently, these two areas of research have interacted with each other under the framework of RL \cite{botvinick2019reinforcement,dabney2020distributional,hassabis2017neuroscience}. This interaction between two areas suggests that RL is a useful approach for studying how humans and AI similarly make decisions in dynamic settings.

Our measure, which we call the \textit{Human-AI Gap}, compares the quality of decisions by humans to the quality of decisions generated by an AI program. Defining the quality of decisions can be a challenge, however, because consequences of decisions are hard to pin down in a high-dimensional state space. Fortunately, modern AI programs based on Deep RL can not only generate decisions of superhuman quality but also evaluate the quality of any decision. Specifically, AI's value network evaluates states, or situations, in a game, allowing us to evaluate how favorable a given state is to the player of interest. Similarly, action-value network evaluates any decision in the given state (producing an output known as the Q value), allowing us to evaluate the quality of any decision in any state. We thus evaluate the quality of any human decision in any state, as well as the quality of a decision generated by AI, and calculate the gap between the two values of decision quality.


\paragraph{Definition of our measure}We use notations from the previous literature \cite{igami2020artificial,silver2017mastering} to explain our measure more formally. It is defined mostly for the environment of the board game Go, but it can be easily modified in other well-defined dynamic decision making problems. State space, $\vert S \vert$, represents a set of possible states of a game. In a match between human players, a human player would face many states in $\vert S \vert$. Given $S$$\in$$\vert S \vert$, a human player would decide the next move. A human player in $k^{th}$ order observes $S_k$ (state) and decides $a_k$ (action), i.e., a position to place the stone. We simplify the decision rule of human players as follows:
\begin{equation*}
    a_k^{Human }=\sigma^{Human}(S_{k};V(S_{k};\theta^{Human}))
\end{equation*}

Human players use their own evaluation parameter ($\theta^{Human}$) to diagnose how advantageous the current state is, $V(S_{k};\theta^{Human})$. Based on the evaluation, they apply their own strategy or decision rule ($\sigma^{Human}$), ending up with an action, $a_k^{Human}$. In this decision rule, we abstract away from complex interactions between human players. Instead, it is more like a single agent problem where each human player has to find an optimal decision in the given state to maximize the total reward. Any strategic responses from the opponent human player are subsumed under the transition of the state in our decision rule.

AI programs also map a given state to an action based on their policy network\footnote{In most cases, Deep Q network is designed to choose an action with the highest expected action-value. In the game of Go, that process is complemented with Monte Carlo Tree Search and AI programs choose a move with the highest playout number. In addition, the dimension of the state space in the board game Go is very large so AI programs take a few key features of each state as input.}. Although the actual process of AI decision making is as complex as human decision making, it can be simplified as follows:

\begin{equation*}
    a_k^{AI}=\sigma^{AI}(S_{k};V(S_{k};\theta^{AI})).
\end{equation*}

It is notable that a human and AI facing the same state, $S_{k}$, may reach a different action (i.e., $a_k^{Human}\neq a_k^{AI}$). This is because the way a human evaluates the state, $\theta^{Human}$, is different from the way AI does, $\theta^{AI}$. Namely, a human may be too optimistic or pessimistic from the perspective of AI. In addition, AI may build a strategy, $\sigma^{AI}$, that does not belong to any traditional human strategy, $\sigma^{Human}$. AI trained by itself is free from any conventional human strategies, so it may produce novel actions. 

We define a measure of the \textit{Human-AI Gap} as follows:
\begin{equation*}
\label{gap}
    \Delta_k \equiv \overbrace{V (S_{k+1}(a_k^{AI}) \;;\theta^{AI})}^\text{Quality of a counterfactual AI decision} -\underbrace{V(S_{k+1}(a_k^{Human})\; ;\theta^{AI})}_\text{Quality of an actual human decision}
\end{equation*}

First, a human player makes a decision, $a_k^{Human}$, after observing a state, $S_k$. We let AI simulate a counterfactual action, which is $a_k^{AI}$, in the same state $S_k$. This action is what AI would choose if it has to decide instead of a human player. Second, we let AI quantify the quality of each action, one from a human player and the other from AI by evaluating the subsequent state. The quality of a counterfactual AI action reflects the current state of the game and acts as a maximum attainable value if a human player chooses the same action as AI. Finally, we calculate the difference between those two evaluations. So our measure quantifies a difference between an advantage induced by a counterfactual AI choice $V (S_{k+1}(a_k^{AI}) \;;\theta^{AI})$, and an advantage induced by an actual human choice $V(S_{k+1}(a_k^{Human})\; ;\theta^{AI})$. If AI comes up with a better action than a human, the \textit{Human-AI Gap} would be positive. If a human chooses the same action as AI, then it would be zero. We use a much superior AI to generate counterfactual actions so this value is usually positive. 
\paragraph{Advantages of our measure} The key advantage of our measure is the ability to place any human decision on a continuous scale of decision quality. Our measure compares the quality of any human decision in any state to a consistent and objective \textit{standard}: the quality of optimal decisions by superhuman AI in the same state. Without this \textit{standard} as a reference point, two human decisions made in two different states could not be compared: just as one decision may be of good quality given a state $S^{'}$, the other decision may also be of good quality given a different state $S^{''}$. How much better one decision is over another will be difficult to determine. However, when the two decisions are evaluated against their respective \textit{standard}, the quality of an optimal decision made by superhuman AI given $S^{'}$ or $S^{''}$, we obtain two values of the gap from their respective AI optima (the \textit{Human-AI Gap}). Using these two values of respective gaps, the quality of the two human decisions can now be put on the same scale and be compared with each other.

Another advantage of our measure is that it leverages more information from AI and is therefore more precise than certain accuracy-based measures in previous research. These accuracy-based measures might calculate the percentage of human decisions that matched AI decisions (e.g., ``share of optimal moves" as in \citealt{strittmatter2020life}) or the percentage of decisions that matched some ground truth (``diagnostic accuracy" as in \citealt{tschandl2020human}). Such measures capture less information than our measure, because they take as their inputs only the discrete, binary values of \textit{whether} human decisions differed from optimal AI decisions or \textit{whether} human or AI decisions deviated from some ground truth. In contrast, our \textit{Human-AI Gap} measure takes as its input \textit{the extent to which} human decision differed from optimal AI decisions on the quality dimension. For example, consider two different human decisions made in the same state $S$, both of which differed from the optimal AI decision in state $S$. An accuracy-based measure would assign an identical value for the two human decisions because they both failed to match the optimal AI decision (e.g., a value of 0, indicating that neither of the two human decisions matched the optimal AI decision). However, our measure would assign different values for the two human decisions, i.e., how much each decision's quality trailed that of the optimal AI decision. Thus, our measure leverages more information from AI than accuracy-based measures and evaluates human decision quality more precisely, on a continuous scale.

Lastly, our measure requires only one type of data: a series of human decisions. As we will show in the next section, we only need human decisions as the sole input to make inferences about human adaptation to AI. Ideally, if we can directly observe the changes in how humans evaluate states ($\theta^{Human}$) or how humans apply rules to make decisions ($\sigma^{Human}$), studying human adaptation to AI would be straightforward. But such direct observation of changes in human decision making parameters would be infeasible. Instead, we can make inferences about whether the parameters have changed by comparing whether and how human decisions have changed in response to the emergence of superhuman AI.

\section{Applications of our measure }
\subsection{Adaptation I: Human learning from AI}
\begin{figure}[ht]
\centering
\begin{subfigure}[ht]{0.47\linewidth}
\centering
\includegraphics[width=4cm]{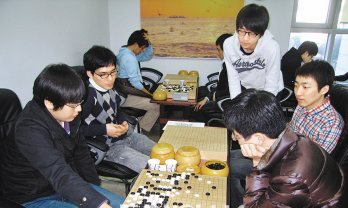}
\caption{Human Learning Before AI}
\end{subfigure}
\hfill
\begin{subfigure}[ht]{0.45\linewidth}
\centering
\includegraphics[width=4cm]{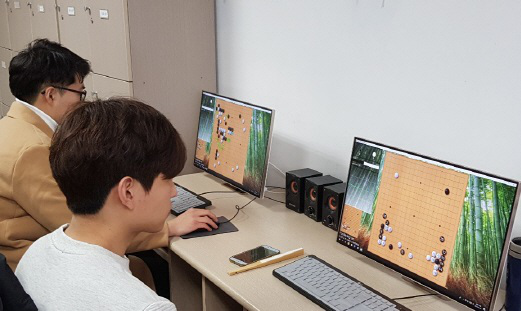}
\caption{Human Learning After AI}
\end{subfigure}%
\caption{Illustration of change of the way human Go players learn winning strategies since AI}
\label{fig:change_human_learning}
\end{figure}

\paragraph{Background} In our first study, we examine the \textit{Human-AI Gap} to compare human players' strategies before and after the launch of AI programs. The measure allows us to evaluate how close each move decision of human players is to the level of AI programs, and thus to determine if human decisions improved after AI programs were available. 
When AI programs did not exist, human players gained new information on Go game strategies by going over top players' move decisions in tournaments, and discussing the strategies with other human players (illustrated in Panel (a)\footnote{``The power of Korean Go? Joint research!" Mar/09/2011 (https://www.donga.com/news/Culture/article/all/20110309/35417308/1)} in Figure \ref{fig:change_human_learning}). What made AlphaGo so sensational in the human Go community is that the AI program used many unorthodox tactics. AlphaGo's demonstration of unfamiliar but effective strategies motivated human players to discuss and learn the strategies of AI. Consequently, the way human players study winning strategies completely changed into getting tutored by AI (illustrated in Panel (b)\footnote{``Hone your Go skills with AI instructors" Feb/26/2019 (http://www.munhwa.com/news/view.html?no=2019022601032103009001)} in Figure \ref{fig:change_human_learning}).

Interestingly, two different AI programs were released at different times. One is AlphaGo in March 2016 and the other is open-source AI programs (e.g., Leela Zero) in October 2017. The main difference between the programs is whether players can observe the programs' intermediate reasoning process behind each move decision. AlphaGo and its subsequent versions show its actions only (i.e., a sequence of positions where AI placed stones). However, the open-source programs and education tools provide information on the detailed thought process behind each AI action. They show the contingency plan of AI such as how AI would respond to a hypothetical state following each counterfactual choice at any states. This allows human players to review the strategies of AI under various states. Also, humans observe how AI evaluates each state of a game (e.g., the current win probability\footnote{Although we use the term \textit{win probability}, its exact interpretation is more subtle. Still it is used in the Go community so we follow this term.} of each player and change in win probability if the human player chooses to deviate from AI's best move). More details on what information is given to human players are explained in the Appendix.
We leverage the fact that two AI programs giving two different learning materials to human players were released sequentially. It gives us a chance to examine the conditions under which human players adapt to AI.

\paragraph{Data}
Our data spans from January 2014 to March 2020 and it comes from official matches between human professional players. It consists of two datasets: (i) 30,995 matches between 357 Korean professional players, where we observe match date, the identity of players, and match outcomes, (ii) 1.3 million move decisions of the Korean professional players from matches between human players. We webscraped the datasets from the Korean Go Association and other websites\footnote{Not every match has been saved with a detailed record of move decisions, but we collected the historical data from multiple sources to get more complete history. The data containing match results spans from 2012 to 2020. The data containing move decisions spans from 2014 to 2020.}. For every single move choice by a human player, we simulate the optimal move decision of AI\footnote{We use an AI program called \textit{Leela Zero} to analyze our data. We use GPU provided by Google Colab Pro (P100) in our simulation.} under the same state of the game and compute our measure by comparing a value network of human players' actual choice and that of AI's choice. Thus, we have 1.3 million move decisions of human players, 1.3 million move decisions of AI, and AI's evaluation of each of these decisions. The gap between a value network of human decisions and that of AI decisions is our measure of the human decision's effectiveness.  

\begin{table}[ht]
\begin{adjustbox}{width=8cm,center}

    \centering 
    \footnotesize{ 
    
    \begin{tabular}{lrrrr}
    \hline
    \multicolumn{5}{l}{Data 1: Player-level match performance (\textit{N} = 357)}\\
    \hline
         & Q1 & Median & Mean & Q3 \\
      Winning rate (\%)   & 31 &46 &43 &81  \\
      \hline
     \multicolumn{5}{l}{Data 2: Move decisions (\textit{N} = 1,357,523)}\\
           \hline
      & Q1 &Median &Mean&Q3  \\
      Move counts within a match   &176&212&216&254\\
      The Human-AI Gap (percentage points)   &0.39&1.68&3.49&4.51  \\
      \hline
    \end{tabular}}
    \end{adjustbox}
    \caption{Summary Statistics}
    \label{tab:summary}
\end{table}

Table \ref{tab:summary} shows the summary statistics of the data. Our data includes heterogeneous players in terms of performance, whose winning rate is 43\% on average. 
For each match, two professional players make an average of 216 move decisions. AI evaluation on each move decision indicates that human players make sub-optimal choices, which results in a 3.5\% loss in win probability. 

\paragraph{Model-free descriptive pattern}

\begin{figure*}[ht]
    \centering
    \includegraphics[width=12cm]{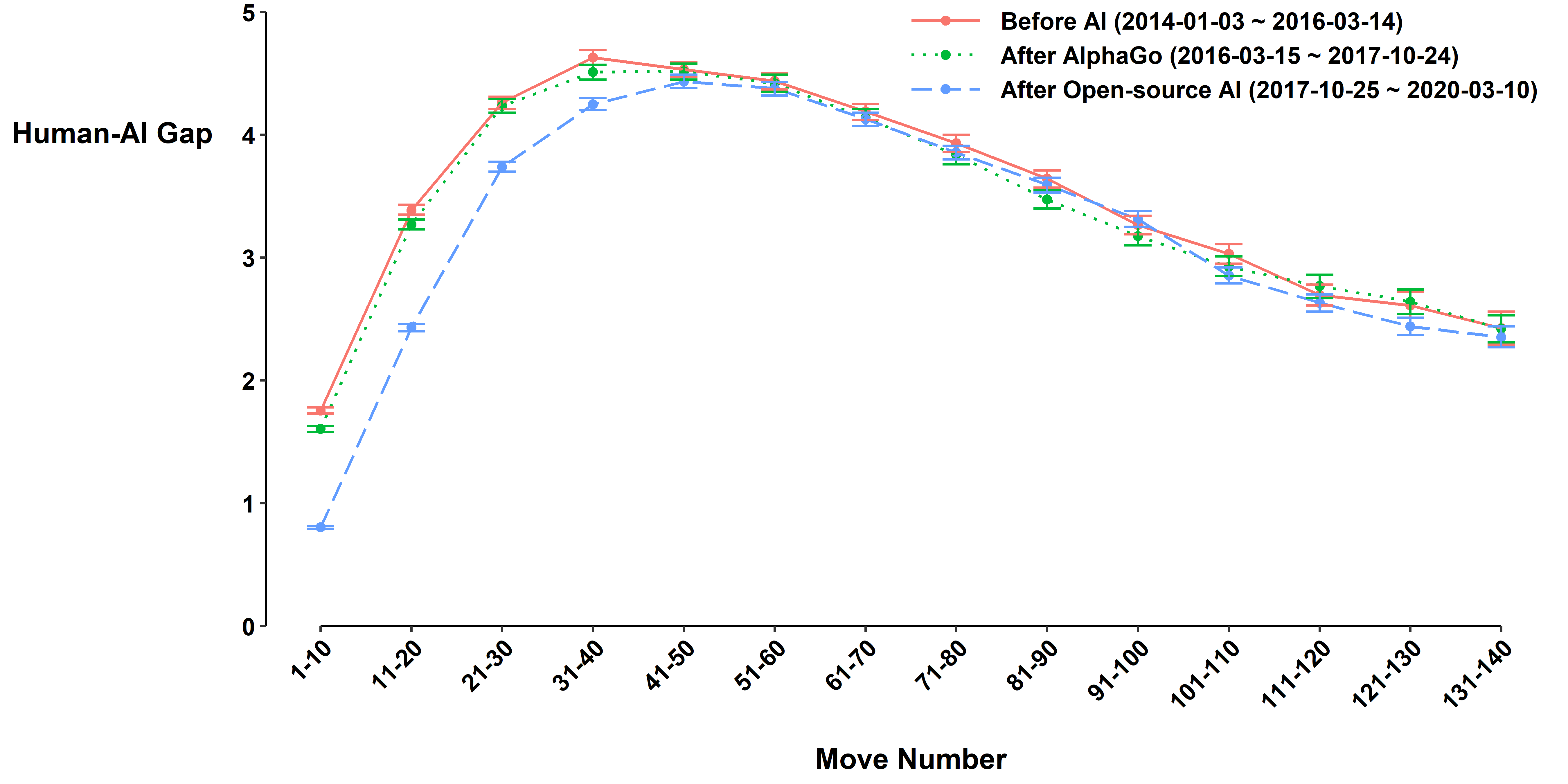}
    \caption{\textbf{Model-free Patterns of the \textit{Human-AI Gap} over the Course of Match (Mean of Each 10 Moves).} The \textit{Human-AI Gap} decreases ever so slightly after AlphaGo's \textit{actions} are released: the dotted green curve is positioned slightly below the solid red curve. After open-source AI programs are released and \textit{reasoning processes} of AI become observable, however, the \textit{Human-AI Gap} decreases by a much greater margin, indicating a marked increase in human decision quality. This decrease in the \textit{Human-AI Gap} occurs mostly for the early game (Moves 1-50). The error bars indicate 95\% confidence intervals around the means for each 10 moves.}
    \label{fig:human_ai_gap_across_moves}
\end{figure*}

We calculate the \textit{Human-AI Gap} ($\Delta$) for every decision made by every human player in our dataset. A value of zero means that a human player completely replicated the AI's decision ($\Delta_k=0 \iff a_k^{Human}=a_k^{AI}$) and made the optimal move, while a positive value indicates the extent to which AI's decision was superior to that of the human player, or put differently, the extent to which the human player's decision quality trailed that of the AI. We present the pattern of $\Delta_k$ in Figure \ref{fig:human_ai_gap_across_moves}. The average \textit{Human-AI Gap} of each set of 10 moves (e.g., $1^{st}-10^{th}$, $11^{th}-20^{th}$, ...) is plotted across the course of match. The solid red curve traces the \textit{Human-AI Gap} before AlphaGo beat Lee Sedol (from January 2014 to March 2016); the dotted green curve traces the \textit{Human-AI Gap} between AlphaGo's debut and the release of open-source AI programs (from March 2016 to October 2017); lastly, the dashed blue curve traces the \textit{Human-AI Gap} after open-source AI programs were publicly released (from October 2017 to March 2020).

\paragraph{Speculation on the inverted U pattern of the \textit{Human-AI Gap} across moves} Perhaps the first thing that jumps out in Figure \ref{fig:human_ai_gap_across_moves} is the inverted U pattern of the \textit{Human-AI Gap} across moves. We speculate that the pattern emerges from two opposing forces at work. First, as the match progresses (and the move number increases), finding the optimal moves becomes harder because stones interact with one another more. For example, in the beginning, when there are only one, two, or several stones, the black and white stones tend not to interact much with each other, as both players seek to control open territory in the corners and sides that are free from either side's influence. In this early stage, human decisions are not too inferior to AI's decisions, leading to a smaller \textit{Human-AI Gap} for earlier moves. As the game progresses (post opening stage), however, black and white stones now clash against each other to fight for territory. In this early to middle stage of the match, optimal moves require thinking not only about how to control more territory as in the beginning, but also how to capture the opponent's stones or how to survive under the opponent's attack. Thus, this clash in the early to middle stage of the match presents a greater challenge for the human decision maker, and as a result, human decisions are inferior to AI's decisions by a greater margin, leading to a greater \textit{Human-AI Gap} for moves in the middle stage. The \textit{Human-AI Gap} peaks at a point and starts decreasing, however, because of the second, opposing force: the more stones there are on the board, the fewer possible moves become available. For example, in the middle stage, there may be 30 possible moves to consider, whereas by the end of the match, there may be only 5 possible moves to consider. With fewer possible moves to consider, the human player can find the optimal moves more frequently, and as a result, human decision quality more closely approaches AI decision quality. We speculate that these two opposing forces give rise to the inverted U shape of the \textit{Human-AI Gap}.

\paragraph{Insights from the pattern of the \textit{Human-AI Gap}} Using our measure of the \textit{Human-AI Gap}, we are able to observe the pattern of human decision quality over the course of a match. One possible insight from this observation may be that room for improvement is greatest in the middle stage of the match, because the middle is where the \textit{Human-AI Gap} is the greatest. Human players may find that studying moves in the middle stage may improve their game more than studying moves in the early or late stage of the match. Thus, our measure of the \textit{Human-AI Gap} can be used to coach human players on what to study. Another possible insight may be that trying to improve human decision making in the middle stage of the match is futile. Human experts may have concluded that improving their middle-stage game is very difficult as compared with improving the early-stage game. That is, although human experts have managed to improve their early-stage game (the red, solid line shifting down to the blue, dashed line for Moves 1-50 in Figure \ref{fig:human_ai_gap_across_moves}), perhaps they could not improve their middle-stage game despite much effort (the blue, dashed line overlaps the red, solid line for Moves 51-140 in Figure \ref{fig:human_ai_gap_across_moves}). Improving the middle-stage game may be very hard, perhaps due to intractable complexity and lack of similarity from one match to another, both of which may prevent discovery or learning of any new principles. If so, human experts may instead double down and focus their effort even more on improving their early- or late-stage game. Whichever insight may reflect reality, our measure could be useful in suggesting how human experts' effort should be allocated.
\begin{figure*}[ht]
    \centering
    \includegraphics[width=11.5cm]{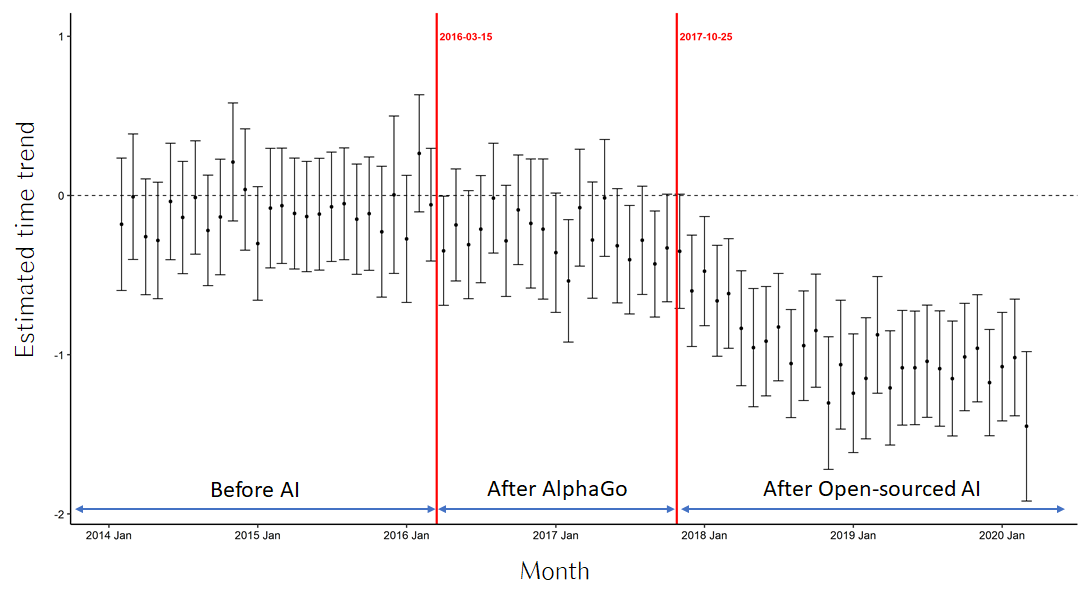}
    \caption{\textbf{Statistical Test on the \textit{Human-AI Gap}} 
    This plot presents the estimated time trend of the \textit{Human-AI Gap} ($\Delta_{it}$) from $\Delta_{it}=\alpha_i+\tau_t+\epsilon_{it}$, where $\alpha_i$ is a player fixed effect, $\tau_t$ is a time fixed effect, and $\epsilon_{it}$ is an error term. The error bars indicate 95\% confidence intervals.
    AlphaGo, despite its superior performance against human, does not help human players to make better decisions. Human players start to make better decisions after the release of the open-source AI. This finding highlights the importance of access to \textit{reasoning process} of AI. }
    \label{fig:human_ai_gap_across_time}
\end{figure*}
\paragraph{Human learning in the early stage of the game} More important than the inverted U pattern are downward shifts in the \textit{Human-AI Gap} (for Moves 1-50). Interestingly, the \textit{Human-AI Gap} decreased only a little bit after human players could observe AlphaGo's actions, as evidenced by a barely noticeable downward shift from the ``Before AI" curve to ``After AlphaGo" curve in Figure \ref{fig:human_ai_gap_across_moves}. In contrast, the \textit{Human-AI Gap} dropped markedly after open-source AI programs became available, as evidenced by a larger downward shift from ``After AlphaGo" curve to ``After open-source AI" curve in Figure \ref{fig:human_ai_gap_across_moves}. Here we do not take into account a difference between players. So we construct a data set at the player-month level ($\Delta_{it}$) in the following way. 

\begin{equation*}
\label{delta_panel}
     \Delta_{it}=\frac{1}{n_{it}}\frac{1}{K}\sum_{j=1}^{n_{it}}\sum_{k=1}^{K}\Delta_{jk}
\end{equation*}
where $\Delta_{it}$ denotes player $i$'s average gap in month $t$. $n_{it}$ stands for the number of matches a player $i$ has in month t, and $k$ represents the order of a choice within a match $j$. Using a panel structure, we can investigate whether human decision quality indeed increased more after open-source AI became available than after AlphaGo's debut. Because the decrease in the \textit{Human-AI Gap} was concentrated in the early and middle stages of the game (Moves 1-50), and because no such decrease was readily observable for Moves 51-140, we focus our attention on the first 50 moves in each match when we investigate human decision quality over time next (so $K=50$).

\paragraph{Insights from the time trend} We examine the \textit{Human-AI Gap} over time after controlling for individual difference. As defined earlier, $\Delta_{it}$ denotes player $i$'s average gap in month $t$. Then we run the following regression: $\Delta_{it}=\alpha_i+\tau_t+\epsilon_{it}$, where $\alpha_i$ is a player fixed effect, $\tau_t$ is a time fixed effect, and $\epsilon_{it}$ is an error term. In Figure \ref{fig:human_ai_gap_across_time}, we report the estimated time trend, $\hat{\tau_t}$. The red vertical line on the left marks March 15, 2016, the date of the match between Lee Sedol and AlphaGo, and the red vertical on the right marks October 25, 2017, the date when the open-source AI program Leela Zero was publicly released, which was followed by releases of similar AI programs and education tools. Consistent with a tiny downward shift in Figure \ref{fig:human_ai_gap_across_moves}, we see little change in the \textit{Human-AI Gap} in the period between the two red vertical lines. The \textit{Human-AI Gap} significantly drops after the second vertical line indicating the time when open-source AI and its analysis tools became available. The decline in the gap over time shows that human players adapted to AI programs gradually, more so after the educational tools were available. The finding provides suggestive evidence that human players modified their choices and did better in Go matches after gaining access to AI.

\subsection{Adaptation II: Cheating from AI}
\begin{figure*}[!ht]
    \centering
    \includegraphics[width = 10cm]{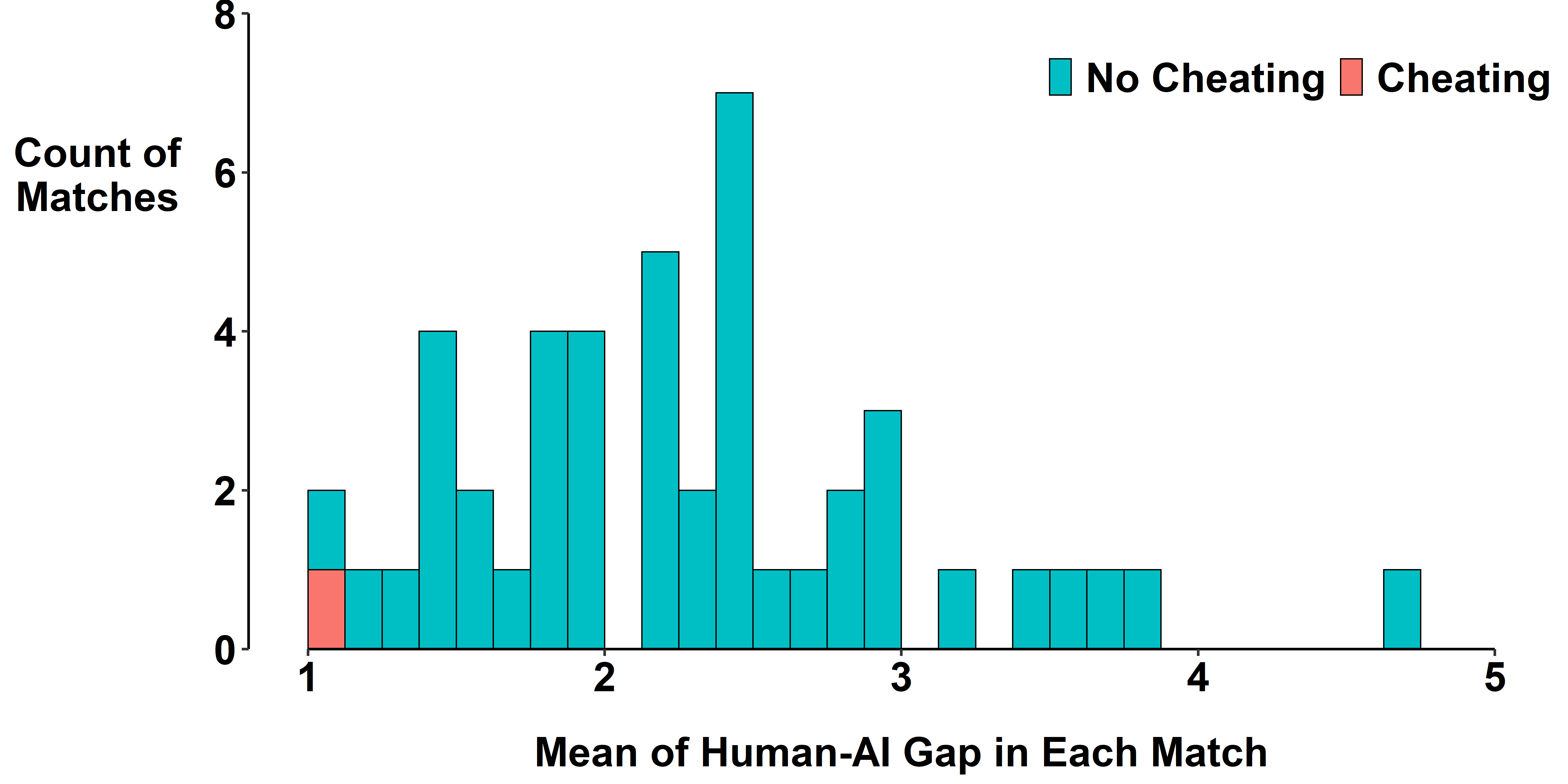}
    \caption{\textbf{A Histogram of Mean \textit{Human-AI Gaps} in the Cheating Player's 52 Career Matches.} Our measure can differentiate the actions where a human player officially committed a cheating by getting help from AI. This study is one example of showing the validity of our measure. In addition, it shows that our measure can be one of possible criteria when examining cheating behaviors of human players.}
    \label{fig:cheating_fig_2}
\end{figure*}
\paragraph{Background}
In our second study, we examine how our measure can be useful in detecting one negative form of adaptation to AI: receiving help from AI to gain an unfair competitive advantage, i.e., cheating. In September 2020, a 13-year-old professional Go player cheated in a match against a 33-year-old top-notch Go player by using an AI program that suggests optimal moves for given states. The high-stakes match (one of Round of 32 matches in a tournament that awards \$175,000 in total prize money) was held online due to a COVID-19 lockdown, opening door to the possibility of cheating, which would have been difficult to pull off had the match been held offline with many eyes around. Not surprisingly, the cheating player showed an extraordinary performance in the match, and as a result, a debate ensued on online forums regarding whether she received help from AI. The controversy culminated in the cheating player finally confessing her transgression in November 2020. Would our measure be sensitive enough to detect such cheating behaviors?

\paragraph{Data}
We first obtain move data for all 52 matches of the cheating player’s professional career, including the match in which she cheated. We then use an AI program to calculate a winning probability associated with each move in these 52 matches and a winning probability associated with an AI’s optimal move. As in the first study, we subtract the former from the latter to calculate the \textit{Human-AI Gap}. To be consistent with the previous study, we focus our analysis on the first 50 moves by each player in each match. This results in 50 values of the \textit{Human-AI Gap} for the cheating player in the cheating match and 2,375 values in the 51 non-cheating matches (after missing values were removed), for a total of 2,425 values in all 52 matches.

\paragraph{Comparison between the cheating match and all other career matches}
We hypothesize that our measure of the \textit{Human-AI Gap} could detect an increase in the cheating player’s decision quality when she received assistance from AI as compared with when she did not. Indeed, that is what we find. In the cheating match, the cheating player’s move decisions show smaller \textit{Human-AI Gaps} (\textit{M} = 1.02, \textit{SD} = 2.13), indicating better decision quality, than move decisions in the non-cheating matches (\textit{M} = 2.35, \textit{SD} = 4.30), \textit{t}(57.78) = 4.23, one-tailed \textit{p} $<$ .001, Cohen's \textit{d} = 0.31 (see Figure \ref{fig:cheating_fig_2}). Two nonparametric tests show converging results. A Wilcoxon rank-sum test reveals that \textit{Human-AI Gaps} are more likely to be smaller when she cheated (\textit{Mdn} = 0.60) than when she did not cheat (\textit{Mdn} = 0.75), \textit{W} = 67883, one-tailed \textit{p} = .041. Similarly, a two-sample Kolmogorov-Smirnov test reveals that the \textit{Human-AI Gaps} in the cheating match and those in the non-cheating matches have different distributions, \textit{D} = 0.23, one-tailed \textit{p} = .006. Results from these three tests show that our measure of the \textit{Human-AI Gap} can be useful to detect higher decision quality from cheating. But can our measure also detect greater \textit{stability} in decision quality from cheating? We hypothesize that the \textit{Human-AI Gap} will be more stable across the moves in the cheating match than in the non-cheating matches, because moves suggested by AI will be consistently optimal, whereas moves made by the human player herself will be optimal less consistently (less frequently). In other words, we want to test whether our measure will exhibit lower variance (more stability in decision quality) in the cheating than in the non-cheating matches. As expected, a Levene's test reveals that variance in decision quality (i.e., the \textit{Human-AI Gap}) is significantly lower in the cheating match (\textit{var} = 4.53, \textit{SD} = 2.13) than in the non-cheating match (\textit{var} = 18.51, \textit{SD} = 4.30), \textit{F}(1, 2423) = 4.85, one-tailed \textit{p} = .014. This study shows that our measure could be useful in detecting a negative form of adaptation to AI, namely receiving AI assistance for an unfair competitive advantage in a professional competition.

\paragraph{Validation of our measure}
In addition to demonstrating the measure's usefulness, this study validates our measure. We know some truth about the world\textemdash{}that cheating occurred\textemdash{}based on the cheating player's public admission of wrongdoing, and our measure is consistent with this truth. In other words, if the \textit{Human-AI Gap} indeed measures adaptation to AI, including cheating, we expect the mean \textit{Human-AI Gap} to be lower in a cheating match than in non-cheating matches, which is exactly what we find. In Section 3 of the Appendix, we similarly validate our measure with data from the first study. Specifically, if the \textit{Human-AI Gap} measures the extent to which human players have learned to play like superhuman AI, then players who won their matches may have played better and more like superhuman AI than players who lost their matches, all else equal. Again, as expected, moves by winning players show smaller \textit{Human-AI Gaps} on average than moves by losing players, which validates our measure.

\section{Discussion}\label{sec:discussion}

\paragraph{Implications} The results from our first study highlight the importance of explainable AI for human learning. Surprisingly, professional Go players did not modify their strategy after observing \textit{actions} of AI (AlphaGo) that beat a top human player. The human experts started making better decisions only after they gained insight into \textit{reasoning process} of AI from open-source AI programs. In Section 2 of the Appendix, we corroborate this finding by comparing the change in decision quality of human players who had access to the AI programs versus who did not\footnote{We examine the \textit{Human-AI Gap} of professional Go players who served in the military when AI programs were available. Those who could not learn from the AI programs due to military service show no improvement over time, providing further evidence to our findings on human adaptation.}. These results suggest that AI designed to improve human decision making will be more effective if it produces outputs that humans can understand themselves rather than outputs that humans must memorize without understanding.

The results from our second study suggest that our measure can assist in cheating detection. The year 2020 was fraught with cheating incidents not only in Go, but also in chess: a grandmaster (ranked No. 260 in the world) was caught cheating in a match against another grandmaster (ranked No. 2 in the world)\footnote{https://www.theguardian.com/sport/2020/oct/16/chesss-cheating-crisis-paranoia-has-become-the-culture}; and Chess.com closed more than 500 accounts \textit{per day} for their ``most common form of cheating, using a chess engine for assistance"\footnote{https://www.chess.com/article/view/online-chess-cheating}. To detect such cheating, Chess.com has developed a system and caught almost half a million cases in its history, including almost 400 titled players and 42 grandmasters. Similar platforms serving chess players or platforms for other games, e.g., poker, may find our measure to be a useful tool or a framework for building a system that guarantees fair play between human players.

\paragraph{Directions for future research} Our measure can shed light on additional ramifications of advance in AI programs in the board game Go. Does the gap between human players increase or decrease? Public release of AI-based analysis tools allow human Go players of \textit{all skill levels} to analyze and simulate consequences of any move at any stage of the game. It is a game changer in how human Go players develop their strategy. In the past, high-ranked players got a competitive advantage to study and discuss latest strategies with other high-rated players. Now low-rated players can learn from AI programs which are much superior to the best human players. But who improves more from the access to AI, weaker or stronger players? On the one hand, weaker players may learn more from AI and catch up to stronger players at a faster rate than they otherwise would have, reducing the performance gap. On the other hand, stronger players may better understand and internalize the baffling yet effective moves by AI algorithm, widening the performance gap. Our measure can be used to answer this question whose implication is broader than the context of the board game Go.

\section{Other potential applications of the measure}

Our measure can be used in contexts where AI programs generate more effective decisions, but humans remain a final decision maker and are held accountable. The key idea of the measure can be used to answer novel questions about human decision making in response to AI. Below, we discuss potential applications of the measure.

Our measure can be used to study factors that lead to slower or faster rate of adaptation to AI. For example, it can reveal that people of certain characteristics (e.g., age, education, or any other relevant background) have advantages or disadvantages in adapting to AI. If such factors are identified, leaders of an organization have to deal with a difficult equity vs. efficiency tradeoff: Should the leadership help the disadvantaged members so that most members of an organization adapt to AI at similar rates and enjoy producing similar levels of output? Or should the leadership focus more on encouraging the advantaged members to maximize adaptation to AI at the organization level? Answering these questions may not be easy, but our measure can nevertheless help identify such factors.

Another natural extension is in the area of personalized education. AI programs leveraging massive amounts of past student data can outperform human teachers in deciding the optimal sequence of materials to present to students at different skill levels \cite{cakmak2012algorithmic,kamalaruban2019interactive}. For example, when teaching students a new set of concepts, teachers may contemplate which concepts to teach earlier and which concepts to teach later, because learning is a dynamic process in which learning in earlier stages affects learning in later stages. 
That is, as in the game of Go, teachers as decision makers must think not only about which concept to teach at given points, but also about how teaching the concept affects learning at later points, all the while trying to maximize overall learning. Teachers can thus learn from AI programs to improve their teaching, and our measure can be used to evaluate teachers' adaptation to AI. 


More broadly, our key idea to compare the output of AI and that of humans is useful in other settings. For example, AI technologies significantly advanced not only in medical diagnosis but also in treatment decisions, such as recommending prescriptions \cite{morley2019debate}. Since each treatment decision would affect the future health status of patients, doctors need to make decisions in a dynamic setting. Even though AI may make better decisions in this context, ultimately human doctors will have the final authority and responsibility in diagnosis and treatment. Recent research shows that collaboration between human doctors and AI significantly improves the accuracy of predictions or diagnoses \cite{tschandl2020human}. As human experts continue with such adaptation to AI, our measure can be used to monitor the progress of the adaptation. Many other decision making problems in business also exhibit features of dynamic optimization \cite{rust2019has}, so firms adopting AI programs as a supporting tool to human managers can use our measure to monitor improvement in human decision making.

\section{Conclusion}
As AI makes better decisions than humans, human experts would adapt to AI. Often, this adaptation comes in a positive form of learning from AI, but it can also appear in a negative form, such as cheating. In this paper, we propose a simple measure, the \textit{Human-AI Gap}, and test whether the measure can detect and quantify human adaptation to AI. Our results using this measure yield valuable insights in the game of Go, such as \textit{when} learning occurred (i.e., after observing AI's reasoning processes rather than its mere actions), \textit{where} learning occurred (i.e., early to middle stage of the match, moves 1-50), \textit{for whom} learning occurred (i.e., experts with access to AI as compared with experts without the access; Section 2 of the Appendix), and \textit{whether} cheating occurred. Moreover, our results suggest that the measure has broader applications in various domains other than game of Go, ranging from managing adaptation to AI in an organization or society, to personalized education, and even to medical diagnosis and treatment.

\section{Acknowledgments}
We appreciate feedback from Kosuke Uetake, Jiwoong Shin, K Sudhir, Elisa Celis, John Rust, Vineet Kumar, Ian Weaver and participants at the 2020 Marketing Science conference and Yale Quantitative Marketing seminar for their valuable and constructive comments. We gratefully acknowledge the support from Korean Go Association to get the data.
\clearpage
\bibliography{references}
\clearpage
\section{Appendix}

\subsection{1. AlphaGo vs. OpenSource AI \& Education Tools}\label{appendix1}
We compare two sets of focal AI programs and discuss their implication from the perspective of human players. 

\paragraph{AlphaGo (including AlphaGoZero or AlphaZero) } 
In March 2016, AlphaGo showed its superhuman performance by beating a world-class human champion. AlpahGo's performance improved even more between 2016 and 2017 and the AI program beat human Go champions in all 64 online Go matches. In May 2017, DeepMind released AlphaGo's moves in 50 matches between AlphaGo and AlphaGo. Then DeepMind made AlphaGo retired. Even though its impact was sensational, the released information to the public was limited to the AI's final actions only.
As summarized in the top plot of Table \ref{tab:two_AI}, human players could have access to AlphaGo's final actions only (i.e., a sequence of positions an AI placed a stone). Because those actions are unconventional, it was not easy for human players to appreciate how AlphaGo would reach its actions. Human players often had heated debates to figure out the rationale behind the decisions of AlphaGo. Although AlphaGo provided new perspectives on the game strategy to human players, it was not the most ideal learning material because it did not provide that rationale to human players.

\paragraph{Opensource AI \& Education Tools} In late 2017, one-and-a-half year after AlphaGo, open-source AI programs such as Leela Zero were released. Unlike AlphaGo of which Deepmind did not publish the source code, the open-source programs enabled developers to devise education tools such as Lizzie. Using the education tool, human players could review their choices in the matches with other human players. They could compare their own evaluation on particular states with the one calculated by AI. By doing so, they could realize whether they were too optimistic or too pessimistic at certain points in the match. In addition, they could observe what AI would have chosen at the state of the match, where they ended up making a sub-optimal move.   

As summarized in the bottom panel of Table \ref{tab:two_AI}, education tools provided information not only on the actions by AI, but also on the detailed thought process behind each action of AI. First, human players observe 
a set of actions AI considers promising as well as the AI's final choice (colored circles in Part 1). The tool also shows how AI would respond to a hypothetical state following each counterfactual choice at the current state (Part 2). Basically, the tool allows human players to review AI's strategy. Here, a strategy refers to a contingency plan of actions under various states. 

Second, human players observe how AI evaluates each state of a game. The number inside colored circles indicates the expected win probability of each action\footnote{It is the output of a Q-function in computer science or a choice-specific value function in economics with the current state and each action as input. Strictly speaking, the technical definition is slightly different from the working definition of win probability but still, it measures the effectiveness of each action.  } (Part 1). Furthermore, the program shows the current win probability of each player (Part 3) and a change in win probability if a human player chooses to deviate from AI's best move (Part 4). For example, human players can try departing from AI's candidate actions (in colored circles), and get feedback on their own choice to quantify any loss the deviation generates. If they disagree with AI choice or evaluation, they can test their strategy and see how AI penalizes it. 

Due to this feature of AI education tools, human players can review their choices in a match with other human players. Altogether, information on AI's strategy and evaluation provided by education tools gives humans an opportunity to understand better AI programs' underlying primitives behind each move decision. 

\subsection{2. Experts with or without the access to AI}\label{appendix2}
\paragraph{Treated and Control players}
Our main strategy to find more convincing evidence is to distinguish human professional players who did and did not have access to AlphaGo or to the subsequent AI programs. We take advantage of the mandatory military service in South Korea. South Korean male citizens are required to serve in the military for 18-24 months before age 29, which forces the military-serving players not to be able to participate in Go matches more than once a month and to be away from recent trends in AI programs related to Go. Specifically, most of them are expected to be confined in their military base but they get 
short-term leave every other month. We confirm from the data that players serving in the military can participate in a tournament once a month at most. They do not have enough time nor a high-performance computer to self-teach unfamiliar tactics, strategies, or insights discovered by AI programs. We have an official record of Korean players' military service history. Thus, a treatment group consists of players who were not serving in the military when AI programs became available, and a control group is players who were serving in the military and did not have access to AI at least for 6 months.

\begin{figure}[htbp]
    \centering
    \includegraphics[width=7cm]{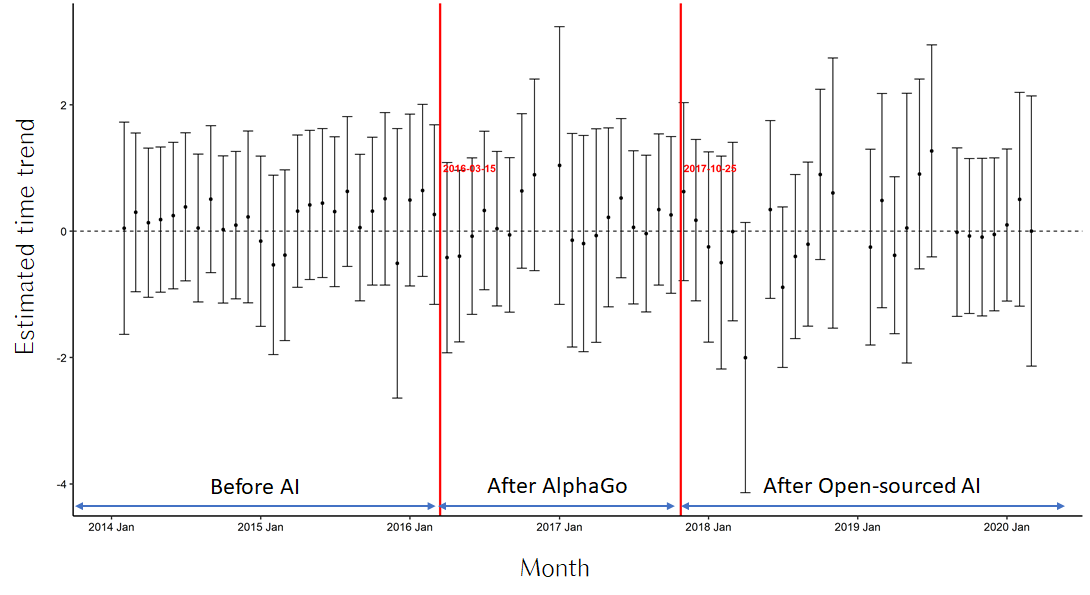}
    \caption{Change in Gap between AI and Humans ($\Delta$) over Time (Control group)} 
    \label{fig:move_plot_comparison_control}
\end{figure}

Figure \ref{fig:move_plot_comparison_control} represents a time trend of $\Delta_{it}$ of players in a control group. In figure \ref{fig:move_plot_comparison_control}, the y-axis is the \textit{Human-AI Gap} ($\Delta$), the x-axis is time, and the two vertical lines represent when the AI programs were available. It shows no significant change over time in our gap measure of control group. It is a sharp contrast to the figure \ref{fig:human_ai_gap_across_time}, which implies that the access to AI programs that make human players choose better move choices. We verify the finding by   Difference-in-Difference estimation. Consider the following regression equation:
\begin{equation*}
\begin{adjustbox}{width=9cm}
\centering
$\Delta_{it}=\alpha_i+\tau_t+\beta\cdot I(i\in \text{Treatment group})\cdot I(t\in \text{Post AI period}) +\epsilon_{it}$
\label{eq_military}
\end{adjustbox}
\end{equation*}

We report the result of $\hat{\beta}$ in table \ref{tab:DiD_estimate}. The dependent variable is the average gap of each player in each month. Each column in the table is based on the different definitions of post-AI periods: all months after Event 1 (March 2016), months between Event 1 and Event 2 (Dec 2017), or months after Event 2 only. The treatment effect of access to AI, captured by the interaction term of (Treatment group) and (Post AI period), is only significant after Event 2. Human players could place their stones in better positions only after they could observe AI programs' strategy (e.g., candidate moves at each state, simulated sequence of actions after the choice) and evaluation (e.g., win probability at each state, change in win probability depending on each candidate move) and understand AI's underlying principles behind each decision. The effect is not significant after they observed AI programs' actions at each state, which might have been considered a ``black-box" that generates incomprehensible outputs. 
\begin{table}[!ht] \centering 
\small
\begin{adjustbox}{width=9.5cm,center}
\begin{tabular}{@{\extracolsep{5pt}}lccc} 
\\[-1.8ex]\hline 
\hline
\\[-1.8ex] &\multicolumn{3}{c}{Definition of \textit{Post AI} periods}\\
\cline{2-4} 
\\[-1.8ex] & After AlphaGo & Between AlphaGo$\sim$open-source AI & After open-source AI\\
\\[-1.8ex] & (March 2016-March 2020) & (March 2016-Oct 2017) & (Oct 2017-March 2020)\\
\hline \\[-1.8ex] 

    (Treatment group) $\times$ (Post AI period)  & $-$0.157 & $-$0.076 & $-$0.278$^{***}$ \\ 
  & (0.099) & (0.133) & (0.094) \\ 
  
\hline \\[-1.8ex] 
Player Fixed Effect &Y&Y&Y\\
Monthly Fixed Effect &Y&Y&Y\\

Observations & 4,887 & 3,026 & 3,435 \\ 
R$^{2}$ & 0.243 & 0.201 & 0.280 \\ 
Adjusted R$^{2}$ & 0.183 & 0.114 & 0.205 \\ 
\hline 
\hline \\[-1.8ex] 
\textit{Note:}  & \multicolumn{3}{r}{$^{*}$p$<$0.1; $^{**}$p$<$0.05; $^{***}$p$<$0.01} \\ 
\multicolumn{4}{r}{Standard errors are clustered at the player level.}\\
\end{tabular} 
\end{adjustbox}
\caption{Diff-in-Diff Estimation of Human Learning} 
  \label{tab:DiD_estimate} 
\end{table} 

To address a concern that serving in the military itself could have affected the \textit{Human-AI Gap}, we study human players who served in the military before AI programs appeared, and implement the same statistical test as before. Specifically, we compare the \textit{Human-AI Gap} between two groups: players that had to serve in the military in 2015 and those that didn't. Then we limit the sample period before March 2016 when AlphaGo came out. We report the coefficient of the interaction term as table \ref{tab:DiD_estimate}. Table \ref{tab:placebo_test} shows that the military serving by itself does not significantly affect the \textit{Human-AI Gap}. Although human players' match performance could have deteriorated, their move decisions were as good as those prior to military-serving according to the \textit{Human-AI Gap}. Thus, we exclude the explanations that the result in \ref{tab:DiD_estimate} is driven by military-serving, and conclude that access to AI drives the effect. 

\begin{table}[!ht] \centering 
\small
\begin{adjustbox}{width=6cm,center}
\begin{tabular}{@{\extracolsep{5pt}}lc} 
\\[-1.8ex]\hline 
\hline \\[-1.8ex] 
 & \multicolumn{1}{c}{\textit{Placebo Test}} \\ 
\cline{2-2} 
\\[-1.8ex] & Before AlphaGo \\ 
\\[-1.8ex] & ($\sim$ March 2016) \\ 
\hline \\[-1.8ex] 
     (Pseudo Treatment group) $\times$ (Pseudo Post period)& 0.081 \\ 
  & (0.090) \\ 
\hline \\[-1.8ex] 
Player Fixed Effect &Y\\
Monthly Fixed Effect &Y\\
Observations & 1,343 \\ 
R$^{2}$ & 0.246 \\ 
Adjusted R$^{2}$ & 0.106 \\ 
\hline 
\hline \\[-1.8ex] 
\textit{Note:}  & \multicolumn{1}{r}{$^{*}$p$<$0.1; $^{**}$p$<$0.05; $^{***}$p$<$0.01} \\ 
\multicolumn{2}{r}{Standard errors are clustered at the player level.}\\
\end{tabular} 
\end{adjustbox}
\caption{Placebo Test} 
  \label{tab:placebo_test} 
\end{table} 

\subsection{3. Validation of Our Measure}\label{appendix3}
\paragraph{The \textit{Human-AI Gap} for Moves by Winning vs. Losing Players}
Does the \textit{Human-AI Gap} tend to be smaller for moves by winning players than for moves by losing players? Answering this question can be one way to test validity of our measure. Our measure would tend to be smaller for moves by winning players who presumably would have played more like superhuman AI than losing players would have. We thus compared the \textit{Human-AI Gaps} of early- to middle-game moves\footnote{We exclude the moves in the very end-game (i.e., Move number $k$ is greater than 150), because the current version of our program uses the Chinese rule for compensating for Black's initial advantage, which creates inaccuracies for calculating the \textit{Human-AI Gaps} for end-game moves within matches under non-Chinese rules. Our data is constructed in matches with non-Chinese rules.} by winning players (\textit{n} = 494,509) to those by losing players (\textit{n} = 494,857) across all matches. The average \textit{Human-AI Gap} was indeed lower\textemdash{}indicating higher decision quality\textemdash{}for moves by winning players (\textit{M} = 3.43, \textit{SD} = 5.15) than for moves by losing players (\textit{M} = 3.88, \textit{SD} = 5.56), \textit{t}(983655) = 41.9, \textit{p} \textless{} .001, \textit{d} = 0.084. A nonparametric, Wilcoxon rank-sum test also showed that the \textit{Human-AI Gaps} were smaller for moves by winning players (\textit{Mdn} = 1.66) than for moves by losing players (\textit{Mdn} = 1.92), \textit{W} = 129616117936, one-tailed \textit{p} \textless{} .001. This indicates that even though the quality of each move is measured from the prediction of self-trained AI, it is also highly correlated with the outcome from matches between human players. 

\clearpage

\begin{table*}[!ht]
  \centering
  \begin{tabular}{ | c | c |}
    \hline
    Event 1: AlphaGo, AlphaGoZero, and AlphaZero & Information to human players  \\ \hline
    \begin{minipage}{.55\textwidth}
    \centering
    \vspace{0.1cm}     
      \includegraphics[scale = 0.08]{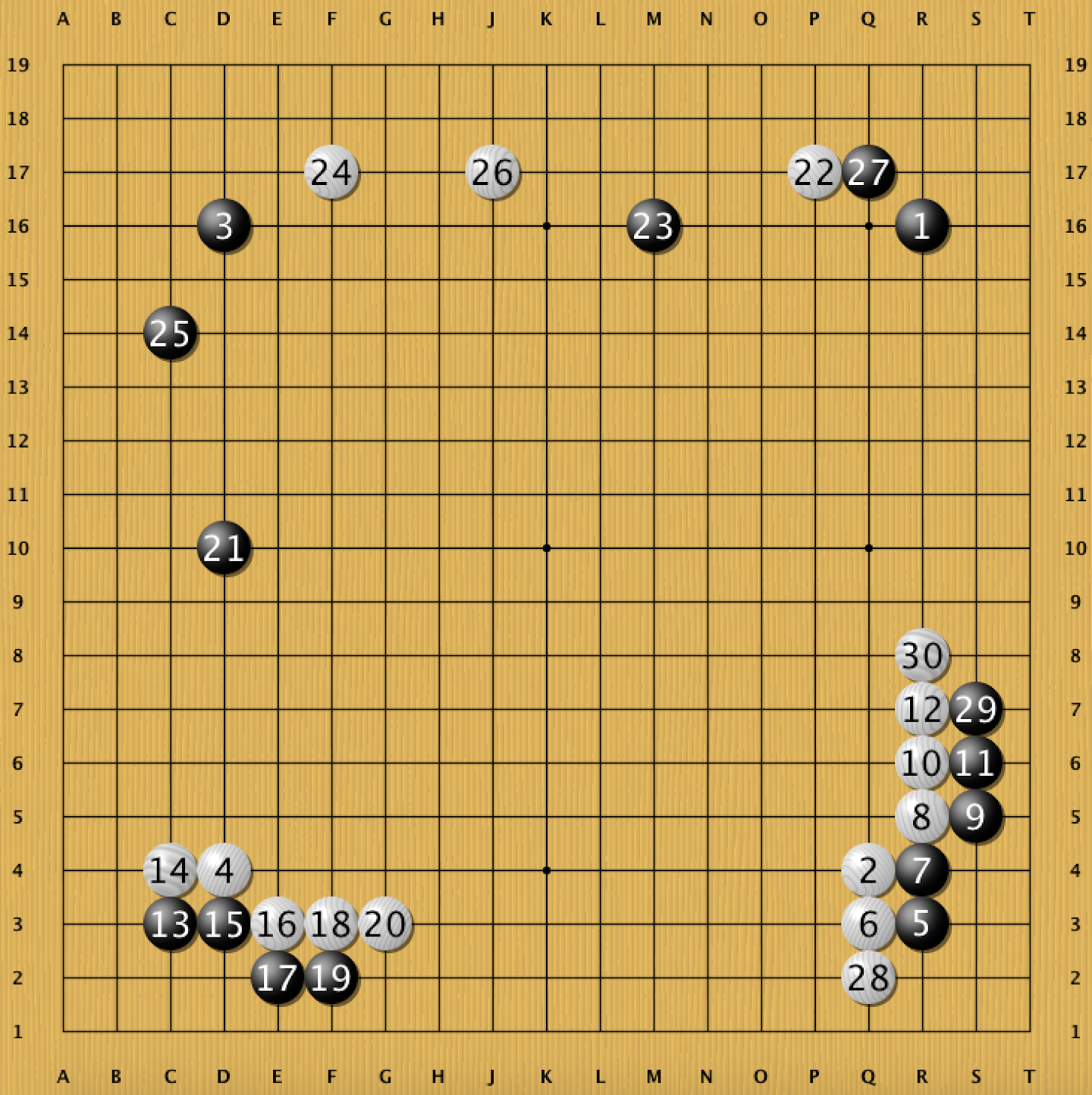}\\
      \vspace{0.1cm}     
    \end{minipage}
    &
    \begin{minipage}{.45\textwidth}
     \begin{flushleft}
    \begin{itemize}
        \item AI's actions only
        \begin{itemize}
            \item The sequence of positions AI placed a stone
        \end{itemize}
        \vspace{0.2cm}
        \item 
         Human players do not observe
        \newline (i) how AI could respond differently to the same situation (No Strategy)
        \newline (ii) how AI predicts consequences of each actual or hypothetical response (No Evaluation)
    \end{itemize}
     \end{flushleft}
    \end{minipage}
    
      
    \\ \hline
    Event 2: Open-source AI \& Analysis Tools  & Information to human players  \\ \hline
    
    \begin{minipage}{.55\textwidth}
    \centering
      \vspace{0.1cm}     
            \small{Part 1. Candidate actions and Win prob of each candidate}
      
      \includegraphics[scale = 0.78]{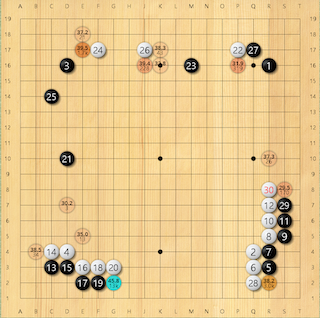} 
      
      \small{Part 2. AI-simulated moves following the choice}
      
      \includegraphics[scale = 0.4]{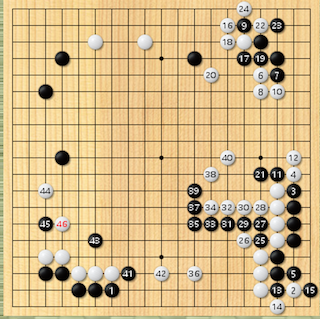}
       
      \small{Part 3. Current win probability of each player}
      
      \includegraphics[scale = 0.2]{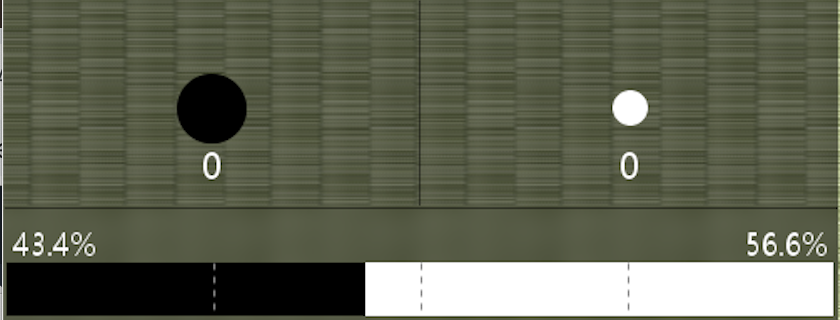}
      
\small{Part 4. Change in win probability over the course of match }

      \includegraphics[scale = 0.25]{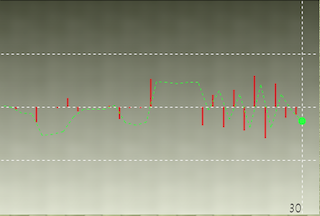} 
    
      \vspace{0.1cm}
    \end{minipage}
    &
    \begin{minipage}{.45\textwidth}
     \begin{flushleft}
    \begin{itemize}
        \item AI's Strategy \& Evaluation, as well as Action 
            \vspace{0.3cm}
    \item 
         Human players observe
         \vspace{0.15cm}
        \newline (i) Strategy: 
        \begin{itemize}
            \item Candidate actions AI considers under the current state (Part 1)
            \item 
            Simulated sequences of actions following the current choice (Part 2)
        \end{itemize}
        
        \vspace{0.2cm}
        (ii) Evaluation:
        \begin{itemize}
            \item Win Probability under the current state (Part 3)
            \item Change in win probability as a consequence of each choice (Part 1)
            \item Change in win probability throughout a match (Part 4)
        \end{itemize}
        
    \end{itemize}
     \end{flushleft}
    \end{minipage}
    \\ \hline
    
\end{tabular}

\caption{AI Programs and its implication in human learning}

  \label{tab:two_AI}
  
\end{table*}

\end{document}